\documentclass[a4paper]{spie}  %>>> use this instead for A4 paper
\usepackage{graphicx}
\usepackage{amssymb}
\usepackage[latin1]{inputenc}
\usepackage[T1]{fontenc}
\usepackage{aeguill}% fontes OK en PDF si pas de package times ou sim.
\newcommand{\dfi}{D_{\phi}}
\newcommand{\ld}{\lambda}
\newcommand{\rapl}{\frac{\lambda_2}{\lambda_1}}
\newcommand{\demi}{\frac{1}{2}}
\newcommand{\va}{\vec{\alpha}}
\newcommand{\vf}{\vec{f}}
\newcommand{\vr}{\vec{r}}
\newcommand{\vro}{\vec{\rho}}

\newcommand{\prior}{\emph{a priori}\ }
\DeclareGraphicsExtensions{.gif,.gif.bb,.ps,.eps,.jpg}

\title{Post processing of differential images  
  for direct extrasolar planet detection from the ground} 

\author{J.-F. Sauvage\supit{a}, L. Mugnier\supit{a}, T. Fusco\supit{a} and
  G. Rousset\supit{a, b}
\skiplinehalf
\supit{a}ONERA, BP-72, 92322 Ch\^{a}tillon Cedex, France; \\
\supit{b}LESIA, Observatoire de Paris, 5 place Jules Janssen, 92195 Meudon, France
}

\authorinfo{email : jean-francois.sauvage@onera.fr}
\begin{document} 
  \maketitle 

%%%%%%%%%%%%%%%%%%%%%%%%%%%%%%%%%%%%%%%%%%%%%%%%%%%%%%%%%%%%% 
\begin{abstract}
  The direct imaging from the ground  of extrasolar planets has become today a
  major  astronomical and  biological  focus. This  kind  of imaging  requires
  simultaneously the use of a  dedicated high performance Adaptive Optics [AO]
  system and a differential imaging camera
  in order to cancel  out the flux coming from the star.  In addition, the use
  of  sophisticated post-processing  techniques  is mandatory  to achieve  the
  ultimate detection performance required. In the
  framework of  the SPHERE project, we  present here the development  of a new
  technique,  based on  Maximum A  Posteriori [MAP]  approach, able  to estimate
  parameters of a faint companion in  the vicinity of a bright star, using the
  multi-wavelength images, the  AO closed-loop data as well  as some knowledge
  on non-common  path and differential aberrations. Simulation  results show a
  $10^{-5}$   detectivity   at  $5\sigma$   for   angular  separation   around
  $15\frac{\lambda}{D}$ with only two images. 
\end{abstract}

%>>>> Include a list of keywords after the abstract 

\keywords{Image processing, exoplanet detection, differential imaging, inverse
problem, regularisation}

%%%%%%%%%%%%%%%%%%%%%%%%%%%%%%%%%%%%%%%%%%%%%%%%%%%%%%%%%%%%%
\section{Introduction}

Today more  than 150 exoplanets have  been detected. But a  great number among
them are  known by indirect gravitational  effects on their  parent star. This
indirect detection and study allows one to estimate physical parameters of the
companion,  like  its  orbital period  or  mass,  but  does not  indicate  its
atmosphere composition or its temperature. Exoplanet direct detection from the
ground represents today a great scientific gain on our knowledge of exoplanet,
since  it  allows one  to  perform  spectroscopy of  the  planet.  But such  a
detection needs a major improvement of technologies in use, since the star and
its companion  are separated by  a fraction of  arcsecond, and the  flux ratio
between   them   is   extremely   high   ($10^6$).   The   SPHERE   instrument
\cite{Beuzit-p-05}, a VLT  Planet Finder, will allow to  detect photons coming
from hot Jupiter planets and will be installed on VLT in 2010. This instrument
is composed  of a  high performance extreme  AO system  \cite{Fusco-p-05a}, an
optimised  coronagraphic   device  \cite{Boccaletti-a-04}  and   a  dual  band
imager\cite{Marois-t-04}. But a  dedicated post-processing method is mandatory
in order to achieve the ultimate  detection level of SPHERE. In this paper, we
will  consider the case  of extreme  AO coupled  to differential  imaging. The
common use of  spectral differential images is to  perform differences between
images  at  different  wavelength  in  order to  calibrate  the  residuals  of
aberration not corrected  by AO and the residuals  of diffraction not canceled
by the  coronagraph. The  main limitation of  differential imaging  comes from
differential  aberrations  between  the  two  images, or  between object images  and
reference images  obtained at  different times.  The  principle of  differential
imaging is  detailed in the next section.  We propose in the  third section an
optimised  method  dedicated  to  our  specific  issue,  based  on  maximum  a
posteriori approach  and able  to estimate the  turbulence parameters  and the
object  in a  pair of  images.  We present  in the  fourth section  simulation
results for the  estimation of the turbulent phase  structure function and the
object.

\section{Spectral differential imaging}

Spectral  differential  imaging  is   an  instrumental  method  that  aims  at
``attenuating''  the flux of  the central  star with  respect to  the flux  of the
potential   companion.   This   method   was   first   initiated   by   Racine
\cite{Racine-a-99}  and Marois\cite{Marois-t-04}.  Thus,  differential imaging
plays  a  role slightly  similar  to  a  coronagraph. The  difference  between
differential imaging  and a coronagraph  is that a coronagraph  subtracts only
the coherent light to the signal, but before detection. Therefore photon noise
in coronagraphic images  is also attenuated. The differential  imaging is able
to subtract also  the star light, but after  detection. The photon noise variance
in differential imaging is therefore  doubled in the combined images. Spectral
differential imaging consists in acquiring two simultaneous images of a system
star-companion  at  different  wavelengths.  These  two  images  are  rescaled
spatially and in intensity and combined in a subtraction that reduces the flux
of  central  star  and of  the  residual  speckle.  Here  we will  only  treat
differential imaging without coronagraph, for simplicity.

The  subtraction  should  reduce  the   star  light,  but  not  the  companion
contribution in  the image. This is  possible if there are  strong features in
the companion spectrum.  In the case of the  giant gaseous exoplanets searched
by  the SPHERE  project, a  strong absorption  band due  to methane  exists at
1.62$\mu m$  and can be  used in such  a subtraction: the  imaging wavelengths
have to be  chosen inside and outside the methane band  (for example 1.575 and
1.625  $\mu  m$), so  that  the  companion emits  at  one  wavelength, but  is
drastically less visible at the other  one. On the other hand, the wavelengths
have to be close enough to ensure the speckle pattern of the central star only
differ in the two images by a spatial and intensity scaling.

Let us  study the  image formation theory  and the limitation  of differential
imaging. The expression  of the two images $i_{\ld_1}$  and $i_{\ld_2}$ can be
written as a convolution of the observed object by the Point Spread Function
[PSF] of the instrument plus  additive noises due to photon statistics and
electronics:

\begin{eqnarray}
  i_{\ld_1}(\va)=h_{\ld_1}(\va) * o_{\ld_1}(\va) + n_1 \nonumber \\
  i_{\ld_2}(\va)=h_{\ld_2}(\va) * o_{\ld_2}(\va) + n_2 
  \label{eqn-images}
\end{eqnarray}

with $h_{\ld_1}$ and  $h_{\ld_2}$ the PSF's in the  two imaging channels which
depends on turbulence  parameters and static aberrations in  the imaging path,
$\va$ the  angular position in  the image field  or the object field,  $o$ the
observed object, *  stands for the convolution process,  $n_1$ and $n_2$ stand
for the  noise in the  images. For unresolved  planets and star,  the observed
object is the  sum of Dirac functions  weighted by the total flux  of the star
and the companions at their respective position. The images are centred on the
star, whose flux is supposed to be the same at the two wavelengths.

\begin{eqnarray}
  o_{\ld_1}(\va) = F_0\delta(\va_0) + \sum_i F_{1,i}\delta(\va_i) \nonumber \\
  o_{\ld_2}(\va) = F_0\delta(\va_0) + \sum_i F_{2,i}\delta(\va_i)
  \label{eqn-obj}
\end{eqnarray}

with  $F_0$ and  $\va_0$ the  total  flux and  position of  the central  star,
$F_{1,i}$,  $F_{2,i}$ and  $\va_i$ the  flux  of the  different companions  at
wavelengths $\ld_1$ and $\ld_2$ and their positions.

There  are  two  different  ways  to perform  image  subtraction:  the  Single
Difference  [SD]  cancels the effect of  the  common  static   aberrations,  the  Double
Difference  [DD] cancels the effect of both  common and  differential aberrations  and is
therefore photon noise limited.

- The SD  consist in directly subtracting the two  images and allows therefore
to cancel  the effect of the  common aberrations in the  two imaging channels.
The images have  to be spatially rescaled at the same  wavelength by a $\rapl$
dilation in the focal plane of the second image.

\begin{equation}
  i_{SD}(\va) \triangleq i_{\ld_1}(\va)-i_{\ld_2}(\rapl \va) 
            \nonumber \\
  \label{eqn-sd}
\end{equation}

If we  consider the  case where  there is no  differential aberration,  and if
$\ld_1$ and  $\ld_2$ are  sufficiently close then  the PSF $h_{\ld_1}$  can be
well approximated by  $h_{\ld_2}$ rescaled at $\ld_1$. The  limitation of this
rescaling  is $\frac{\Delta  \ld}{\ld}$ \cite{Marois-t-04}.  Therefore  the SD
gives a  good approximation of the  difference of the  companions convolved by
first PSF $h_{\ld_1}$, as the star light has been totally reduced:

\begin{equation}
  i_{SD}(\va)=\sum_i \left(F_{1,i}\delta(\va_i)-\sum_i F_{2,i}\delta(\rapl
    \va_i)\right) * h_{\ld_1} + n_1-n_2
            \nonumber \\
  \label{eqn-perfect_case}
\end{equation}

But in a more realistic case, the static differential aberrations are not null
and the  difference between  the image $i_{\ld_1}$  and the  image $i_{\ld_2}$
rescaled at $\ld_1$  makes appear the effect of  differential aberrations.

The two  images have to  be acquired simultaneously  in order to see  the same
acquisition  conditions  (turbulence  parameters,  guide  star  magnitude,  AO
performance...).  Two  imaging  channels  are  therefore used,  each  of  them
acquiring an image centred on the imaging wavelength. The efficiency of this
subtraction  depends  on differential  aberration amplitude  between  the two  optical
imaging channels, since these aberrations are the main difference between the
two combined images \cite{Boccaletti-p-05b}.

- The DD aims at solving the SD limitation by using  two reference images obtained
on a reference star with the same  imaging tool but at another time, the DD
therefore cancels the effect of the differential aberrations (assuming that
they have not evolved between the two observations):

\begin{equation}
  i_{DD}(\va)=\left( i_{\ld_1}(\va)-i_{\ld_2}\left(\rapl \va\right) \right)
             -\left( i_{ref, \ld_1}(\va)-i_{ref, \ld_2}\left(\rapl \va\right) \right)
  \label{eqn-DD}
\end{equation}

The reference  images are acquired at  a different time, and  on a different
position on  sky. This method is  therefore sensitive to the  evolution of the
observing conditions between  the acquisition of the  two pairs of  scientific and reference
images. The  evolution of turbulence  parameters, AO performance, and  most of
all the evolution of quasi-static  aberrations are the main limitations of the
DD method.

\section{Post processing for differential imaging}

As explained  before, the  detection of low  flux companions  (contrast around
$10^6$ between central star and companion) requires the perfect calibration of
both differential  static aberrations and system  parameters (AO performance).
In  a first  approximation,  we assume  that  the static  aberrations in  each
imaging channel  are perfectly known. This  is well achieved by  using a phase
diversity calibration,  as described by Sauvage  et al.\cite{Sauvage-p-05}. In
this  framework, we present  here a  new post-processing  deconvolution method
based  on a MAP  approach that  estimates the  turbulence-induced PSF  and the
observed object.

\subsection{Separation static / turbulent aberrations in long exposure images}

The image  formation from the  ground of stellar  objects is perturbed  by two
factors:  the  atmospheric  turbulence  and  the  static  aberrations  of  the
telescope. The aberrant pupil phase is therefore the sum of two terms: $\phi =
\phi_t + \phi_s$ with $\phi_t$ the turbulent part and $\phi_s$ the static part
of the  phase. The turbulent phase $\phi_t$  is a random variable  of time and
position  in pupil  plane  and  is therefore  characterised  by its  structure
function $\dfi$, whereas the static phase $\phi_s$ does not depend on time and
is     deterministically    known.     If    the     turbulent     phase    is
stationary\cite{Conan-t-94} (as  for uncorrected turbulence) then  it has been
shown by  Roddier \cite{Roddier-81a} that the  OTF is the product  of the long
exposure turbulence-induced OTF and of the static OTF:

\begin{equation}
  \tilde{h}(\vf)=\exp\left(-\frac{1}{2}D_{\phi}(\ld
  \vf) \right) \frac{1}{S_{pup}} \int\!\!\!\!\int_{S_{pup}} P(\vec
  r + \ld \vf)\exp\left(i.\phi_s(\vr + \ld \vf)\right).P(\vec r)^*.\exp\left(-i.\phi_s(\lambda
  \vf)\right) d^2\vec r \nonumber \\
  \label{eqn-sepa}
\end{equation}

with

\begin{itemize}
\item $P(\vec r)$ the pupil function
\item $D_{\phi}(\lambda \vf)$ the atmospheric phase structure function
  after AO correction at wavelength $\lambda$: 
  \begin{equation}
    D_{\phi_t}(\vro)  \triangleq \langle|\phi_t(\vec{r} + \vro)-\phi_t(\vec{r})|^2\rangle
  \end{equation}
\end{itemize}

The  phase structure function  $D_{\phi_t}(\lambda \vec  f)$ is  a statistical
term that quantifies  the turbulent phase variations for  two points separated
by $\vro=\lambda  \vec f$ in  the pupil plane and its shape depends on
turbulence parameters and on AO performance. If the turbulence is corrected,
$\dfi$ depends both on $\vr$ and $\vro$

\begin{equation}
  D_{\phi_t}(\vr, \vro) = \langle|\phi_t(\vec{r} + \vro)-\phi_t(\vec{r})|^2\rangle
  \label{eqn-dphi}
\end{equation}

The  average $\langle  \cdot \rangle$  in the  expression  of $D_{\phi_t}(\vr,
\vro)$ is  theoretically an average on  phase occurrences (and  thus on time),
but may  be approximated  by an average  $\langle \cdot  \rangle_{\vec{r}}$ on
$\vec{r}$   (stationarity  approximation). This  simplified expression
$D_{\phi_t}(\vro)=\langle|\phi_t(\vec{r}                                      +
\vro)-\phi_t(\vec{r})|^2\rangle_{\vr}$ is  therefore independent of $\vec{r}$.
This stationarity  approximation is  justifiable in the  case of  a Kolmogorov
turbulence  statistic,  and  often  used  also in  the  case  of  AO-corrected
turbulent phases\cite{Conan-t-94}.

Equation~(\ref{eqn-sepa})  shows that  the  global  OTF is  the  product of  a
turbulence-induced OTF and a static OTF:

\begin{equation}
  \tilde{h}(\vf)=\tilde{h}_t(\vf) \cdot \tilde{h}_s(\vf)
  \label{eqn-sepa_h}
\end{equation}

with $\tilde{h}_t(\vf)$ the long exposure OTF due to turbulence only, and
$\tilde{h}_s(\vf)$ the OTF due to telescope and the aberrations.

The  structure function at  a wavelength  $\ld_2$ can  be rescaled  at another
wavelength $\ld_1$  by the operation  described in Equation~\ref{eqn-rescale},
in order  to compute the turbulence-induced  OTF in the first  image. Thus, the
turbulent OTF $\tilde  h_{t,\ld_1}$ in the first image  can be computed thanks
to this structure function  at $\ld_2$ by using relation~\ref{eqn-otf_image2}.
The  image  formation  described   in  Equation~\ref{eqn-images}  can  now  be
rewritten as  in Equation~\ref{eqn-images_bis} taking  explicitly into account
the turbulent and static components of the phase.

\begin{equation}
  D_{\phi_t, \ld_1}(\vro) = (\rapl)^2 D_{\phi_t, \ld_2}(\rapl \vro)
\label{eqn-rescale}
\end{equation}

\begin{equation}
  \tilde h_{t,\ld_1} = \exp\left(-\demi (\rapl)^2 D_{\phi_t,\ld_2}(\rapl \vro)\right)
  \label{eqn-otf_image2}
\end{equation}

\begin{eqnarray}
  i_{\ld_1}(\va) &=& h_{t,\ld_1}(\va) * h_{s,\ld_1}(\va) * o_{\ld_1}(\va) \nonumber \\
  i_{\ld_2}(\va) &=& h_{t,\ld_2}(\va) * h_{s,\ld_2}(\va) * o_{\ld_2}(\va)
  \label{eqn-images_bis}
\end{eqnarray}

with $h_{t,\ld_1}$ and $h_{t,\ld_2}$ the turbulent long exposure PSF depending
on  turbulent  phase structure  function  $D_{\phi,\ld}(\vro)$, $h_{s,\ld_1}$  and
$h_{s,\ld_2}$   the  PSF   depending  on   static  aberrations   $\phi_{s,1}$  and
$\phi_{s,2}$ in the two imaging channels.

\subsection{The post-processing framework}

The  main   limitation  of   differential  imaging  comes   from  differential
aberrations in the  two spectral channels which creates different  static pattern in the
images in the  case of SD, and  the evolution of these patterns  due to system
state modification in the case of DD. Therefore in our new approach we propose
to estimate the PSF and the  observed object $o$ in the two images $i_{\ld_1}$
and $i_{\ld_2}$.  The estimation of  the PSF reduces  to that of  its residual
turbulent  component as  the static  aberrations are  supposed to  be measured
separately. The estimation  is done thanks to the  minimisation of an adequate
MAP criterium $J(\dfi,o)$.

The  MAP  approach  is  based   on  writing  the  probability  $\mathcal{P}  =
P(o,\dfi|i_{\ld_1},i_{\ld_2})$ of a given object and $\dfi$ knowing the images
using Bayes'  theorem (see Equation \ref{eqn-proba}). Finding  the best object
and  structure function  means maximising  the probability  $\mathcal{P}$ with
respect to $o$ and $\dfi$.

\begin{equation}
  \mathcal{P}(o,\dfi) = P(o,\dfi|i_{\ld_1},i_{\ld_2}) \propto 
  P(i_{\ld_1},i_{\ld_2}|o,\dfi) \cdot P(o) \cdot 
  P(\dfi)
  \label{eqn-proba}
\end{equation}

The first factor $P(i_{\ld_1},i_{\ld_2}|o,\dfi)$ is called ``likelihood term''
and embodies  the  relationship  between  data  and  the  sought  parameters.  Its
statistics is  given by  the noise statistics  in the image  (stationary white
Gaussian  noise in  a first  approximation). The  other probabilities  are the
\prior knowledge we  have on the parameters to  estimate. These regularisation
terms allow  to smooth the criterium  and to accelerate  its minimisation. For
instance,  the  turbulent phase  structure  function  has  a particular  shape
depending on turbulence parameters and  may therefore be taken into account in
this  regularisation term.

The  criterium   to  minimise   is  $\mathcal{J}  =   -\ln(\mathcal{P})$  with
$\mathcal{P}$  written in  Fourier space  and  may be  rewritten as  (Equation
\ref{eqn-J}). 

\begin{eqnarray}
\mathcal{J}(\dfi,o)&=&
  ||\tilde i_{\ld_1}(\vf)-\tilde h_{t,\ld_1}(\dfi,\vf)\cdot\tilde
  h_{s,\ld_1}(\vf)\cdot\tilde o_{\ld_1}(\vf)||^2 
  \nonumber \\
  &+& ||\tilde i_{\ld_2}(\vf)-\tilde h_{t,\ld_2}(\dfi,\vf)\cdot\tilde
  h_{s,\ld_2}(\vf)\cdot\tilde o_{\ld_2}(\vf)||^2 
  \nonumber \\
  &+& \mathcal{J}_{R,\dfi}(\dfi)+ \mathcal{J}_{R,o}(o)
  \label{eqn-J}
\end{eqnarray}

where $\tilde{  }$ denotes  the Fourier transform,  $\mathcal{J}_{R,\dfi}$ and
$\mathcal{J}_{R,o}$   denote  regularisation   terms  accounting   for  \prior
knowledge we may have on the parameters to estimate.

\subsection{Assumption and subsequent simplified method}

In  the  framework  of  this  deconvolution process,  we  make  the  following
assumption  in  order  to   simplify  the  minimisation  and  demonstrate  the
feasibility of such a global technique: 

We assume that the companion presents particular spectral signature : it emits
light  at  the  first  wavelength  and  is totally  undetectable  at  the  second
wavelength, it  means the object is  an ideal hot Jupiter  and presents strong
absorption line  around 1.6$\mu$m. The second image  $i_{\ld_2}$ can therefore
be seen as a calibration PSF,  and the global minimisation may be approximated
by the three following steps :

\begin{itemize}
\item[1)] Estimation of the  structure function $\dfi(\ld_2 \cdot \vec{f})$ in
  the  image $i_{\ld_2}$ without  the object,  knowing the  static aberration.
  This corresponds to the minimisation of  the two middle term with respect to
  $\dfi$  in   Equation~\ref{eqn-J}:  likelihood  term   on  $i_{\ld_2}$,  and
  regularisation term on $\dfi$.
\item[2)]  Rescaling  of the  structure  function  (estimated  at $\ld_2$)  at
  wavelength $\ld_1$  according to Equation~\ref{eqn-rescale}  and computation
  of  global PSF  $h_1$ of  the first  image $i_{\ld_1}$,  knowing  the static
  aberration of the first channel according to Equation~\ref{eqn-otf_image2}.
\item[3)] Deconvolution  of the first  image with the previously  computed PSF
  $h_1$, and estimation of the  object in $i_{\ld_1}$. This corresponds to the
  minimisation of the two terms depending  on $o$ only (first and last) in the
  criterium  $\mathcal{J}$ with respect to $o$ .
\end{itemize}

\subsection{Estimation of phase structure function  $\dfi$}

\begin{figure}
\begin{center}\leavevmode
  \includegraphics[width=.7\linewidth]{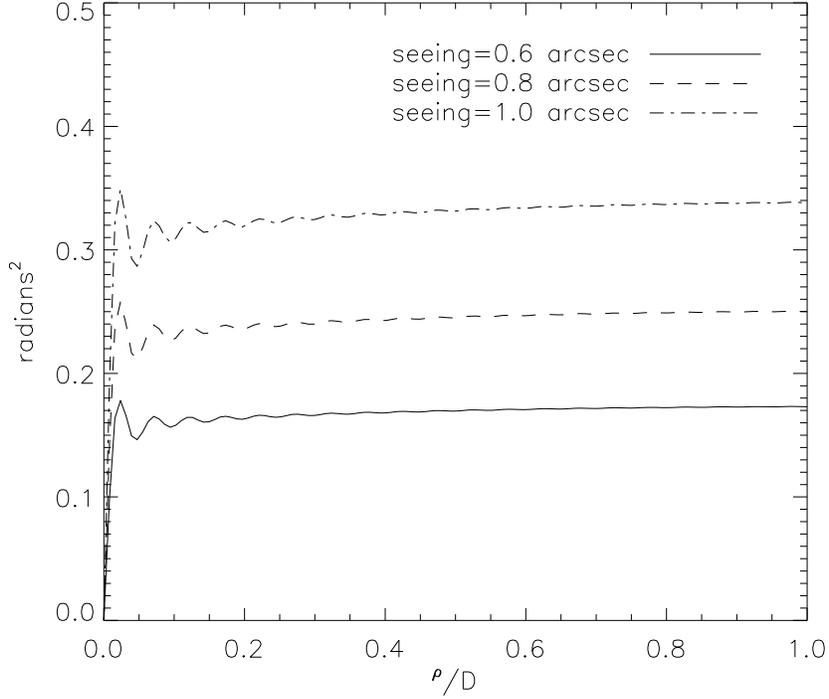}
  \caption{Profiles  of  phase  structure   function  for  a  turbulent  phase
    corrected by AO. Condition of  simulation : Paranal + SAXO, with different
    values of seeing.}
  \label{fig-dphi_prof}
  \end{center}
\end{figure}

The  turbulent phase  structure function  gives a  statistical knowledge  on a
turbulent phase.  For a  turbulent phase following  a Kolmogorov  profile, the
structure       function       is       given      by       the       relation
$\dfi(\rho)=(\frac{\rho}{r_0})^{\frac{5}{3}}$, with  $r_0$ the Fried parameter.
But for a turbulent phase corrected  by an AO system, this relation takes into
account the AO system parameters and is here numerically estimated. The Figure
\ref{fig-dphi_prof} shows typical profiles of $\dfi$ for the turbulence and AO
conditions explained in section \ref{sec-results}, with variations of seeing.

Let  us study  the first  step of  the method  : the  estimation  of structure
function  $\dfi$ in a  calibration long  exposure   PSF. A  criterium (see
Equation  \ref{eqn-crit_dphi}) is  used for  this minimisation,  based  on the
likelihood term and a regularisation term on $\dfi$.

\begin{eqnarray}
  \mathcal{J}(\dfi)&=&||\tilde i_{\ld_2} - F \cdot \exp(-\demi \dfi) \cdot \tilde
  h_{s,\ld_2}||^2 \nonumber  \\
  &+& \mathcal{J}_{\dfi}(\dfi)
  \label{eqn-crit_dphi}
\end{eqnarray}

with  $F$ the  flux of  the observed  star, $\dfi$  the structure  function to
estimate and $\tilde h_{s,\ld_2}$ the  static PSF due to static aberration. $\dfi$
is the  only estimated parameter,  since the star  flux as well as  the static
aberrations (and therefore the static PSF) are assumed to be known.

The regularisation  term $\mathcal{J}_{\dfi}(\dfi)$ is  an adaptive smoothness
term  on  estimated  $\dfi$  designed  to  avoid  noise  amplification  during
estimation  and to  allow the  extrapolation of  $\dfi$ to  regions  where the
static OTF is very small (or  even null). The regularisation is done using the
gradient of structure function  $\nabla \dfi$ and penalises deviations between
two adjacent  pixels according  to a typical  adaptive variance,  depending on
pixel   position.   This  term   is   computed   as   explained  in   Equation
\ref{eqn-regul}.

\begin{equation}
  \mathcal{J}_{\dfi}(\dfi) = \demi \left(\nabla
  \dfi\right)^t C_{\nabla \dfi}^{-1} \left(\nabla \dfi\right)
  \label{eqn-regul}
\end{equation}

with  $C_{\nabla  \dfi}$  the  covariance  matrix of  the  gradient  of  phase
structure function $\nabla \dfi$. This covariance matrix has been estimated on
different occurences  of $\nabla \dfi$,  these occurences have  been generated
with  different value  for $r_0$,  wind  speed or  star magnitude.  $C_{\nabla
  \dfi}$ quantifies the typical variability of $\nabla \dfi$ and allows one to
correctly  weigth  the regularisation  term.  A  common  issue in  regularised
inversion   methods  and  criterium   minimisation  is   how  to   choose  the
hyperparameter,  that  balances the  two  terms  of  the criterium.  With  the
Bayesian  apporach  adapted here  and  with the  use  of  a $C_{\nabla  \dfi}$
estimated by simulations, there is no such hyper-parameter to be tuned and the
estimation of $\dfi$ is completely  unsupervised.

\subsection{Object estimation}

In our  procedure, the structure function  and the object  estimation are done
sequentially.  This  simplified approach  gives  a  good  idea of  the  global
approach performance,  even though a  global minimisation should be  even more
precise and  therefore lead to a  slightly better object  estimation (which is
the final goal).

The  object  estimation  is  done using  MISTRAL\cite{Mugnier-a-04}  algorithm
developed  at  ONERA. This  algorithm  is based  on  the  minimisation of  the
following criterium,  and gives the  best object given  an image, its  PSF and
\prior knowledge:

\begin{equation}
  \mathcal{J}(o) = ||i_{\ld_1}-\hat h_{t,\ld_1}*h_{s,\ld_1}*o||^2 + \mathcal{J}_R(o)
  \label{eqn-mistral_crit}
\end{equation}

with  $\hat  h_{t,\ld_1}$ the  turbulent  PSF  at  $\ld_1$ computed  with  the
estimated  $\dfi$,  $\mathcal{J}_R(o)$ a  regularisation  term accounting  for
\prior  knowledge  on  the  object.  This  regularisation  term  may  contains
different terms.  In our particular problematic, we used
a positivity constraint and a quadratic linear-quadratic regularisation\cite{Mugnier-a-04}.

\section{Results}
\label{sec-results}
In this section, we validate our post-processing method on simulated data. The
simulation conditions are detailed in  the following list, and correspond to a
8m class  Telescope with  an Adaptive Optics  system of high  performance like
SPHERE, and  a turbulence  profile corresponding to  a typical Paranal  sky. The
goal of  this simulation is to  compare the detectivity  of Single Difference,
Double Difference and our approach.

Conditions :
\begin {itemize}
\item  $\ld_1=1.60\mu m$,  $\ld_2=1.58\mu m$  (corresponding to  two wavelengths
  inside and outside the methane absorption line)
\item Turbulence parameter  : a typical $Cn^2$ profile for Paranal is being
  used with an average wind speed of 12.5 m/s, and seeing of 0.8 arcsecondes.
\item Adaptive Optics parameters  : as extreme-AO. 41$\times$41 actuators with
  spatially filtered  Shack Hartmann WFS,  a L3CCD working at  1.2kHz sampling
  frequency. The guide star has a V-magnitude of 8.
\item The  static aberration  component is randomly  generated according  to a
  $\frac{1}{n^2}$  spectrum ($n$ being  the radial  Zernike order)  with 1300
  Zernike coefficients and a differential  wavefront error of 10nm RMS in
  each channel. i.e., the total differential wavefront error is 14nm RMS.  
\item Imaging parameters : 256$\times$256  images, with a 8m telescope. The different
  PSF's   $h_{\ld_i}$   are   generated   at   Shannon  (i.e.,   one   pixel   is
  $\frac{\ld_i}{2D}$  arcsecond on  sky) and  are therefore  already spatially
  rescaled.  The   images  $i_{\ld_1}$   and  $i_{\ld_2}$  are   generated  by
  convolution of the object and the PSF of each channel.
\item The object at the first wavelength  is a star and three companions with
  a flux
  ratio of $10^{-3}$ for the first image, and the star alone for the object at
  the second wavelength. The star flux is set to a total of 10$^7$ photons for
  each wavelength. The companions are located close to the star, respectively
  at 2.5, 5.0 and 7.5 $\frac{\ld}{D}$.
\end{itemize}

\begin{figure}[h!]
\begin{center}\leavevmode
  \includegraphics[width=.7\linewidth]{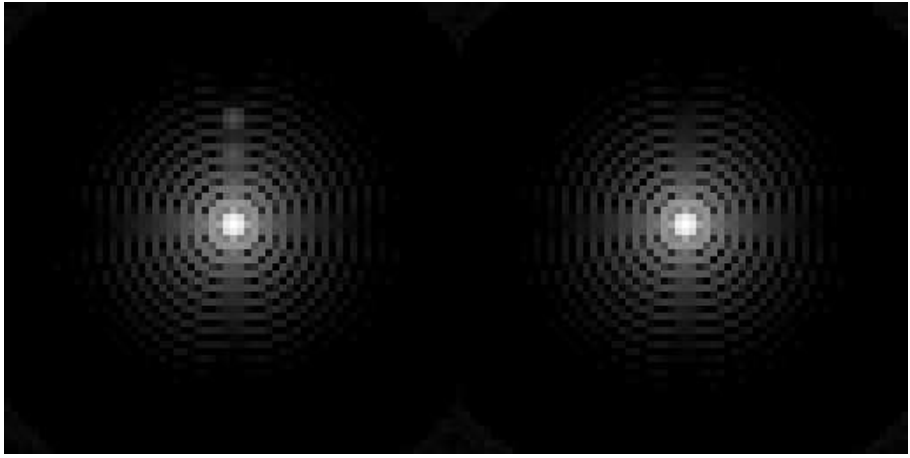}
  \caption{Example of spectral   images, logarithmic scale, two of
    the three companions around the central star are visible on the left image
    ($\ld_1=1.6\mu m$), and  only the star in the  right image ($\ld_2=1.58\mu
    m$).}
  \label{fig-images}
  \end{center}
\end{figure}

Such conditions  allow us to generate  quite realistic images  (see example on
Figure \ref{fig-images}) that are processed by our method.

\subsection{$\dfi$ estimation : simulation results}

\begin{figure}
\begin{center}\leavevmode
  \includegraphics[width=.7\linewidth]{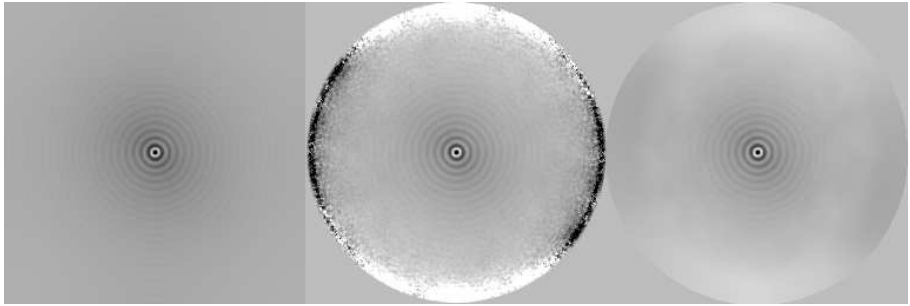}
  \caption{True $\dfi$, $\dfi$ estimated without regularisation, 
    $\dfi$ estimated with regularisation.}
  \label{fig-dphi_est}
  \end{center}
\end{figure}

\begin{figure}[h!]
\begin{center}\leavevmode
  \includegraphics[width=\linewidth]{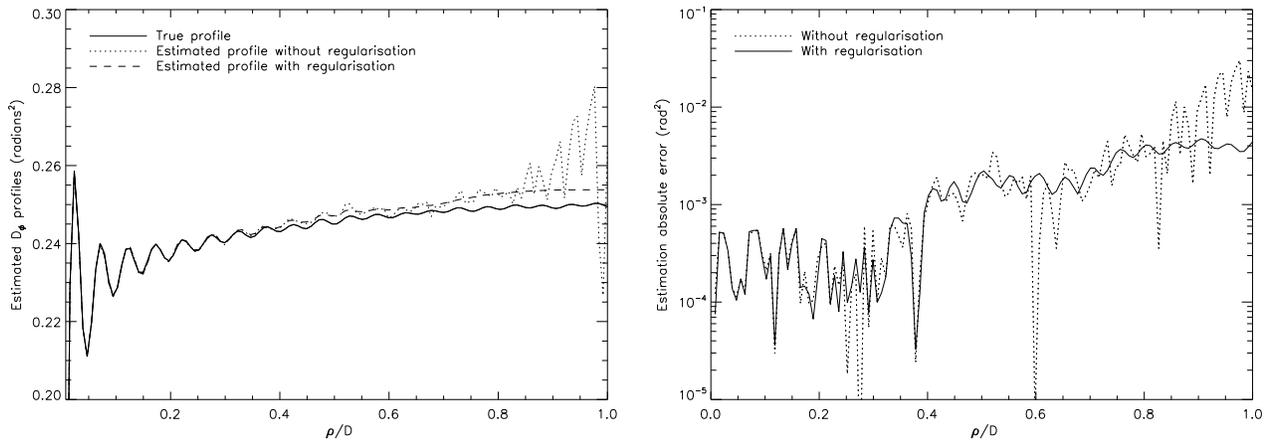}
  \caption{X cross-section  for [left]  true and estimated  structure function
    with and  without regularisation, and [right] absolute  error on estimated
    structure function with and withoutu regularisation.}
  \label{fig-profiles}
  \end{center}
\end{figure}

The image  $i_{\ld_2}$ is processed in  order to estimate  the phase structure
function via the minimisation of  the criterium presented in previous section.
Figure~\ref{fig-dphi_est} shows results of  $\dfi$ estimation. The true $\dfi$
(used to generate the images) on  the left shows the plateau value and central
features characteristic of  AO system. In the middle,  the estimated structure
function  without regularisation  (only the  likelihood  term is  used in  the
criterium). Noise on the edge of the circular support of $\dfi$ is amplified. The
use of adaptive regularisation (on the  right) allows us to reduce this noise
amplification and gives a far  better estimation of $\dfi$. The error profiles
are  plotted  on  Figure   \ref{fig-profiles}.  

Without  regularisation,  the  error   is  lower  than  0.03  rad$^2$.  Figure
\ref{fig-profiles}  shows  the  gain   brought  by  regularisation  on  $\dfi$
estimation : the maximum error for  high frequencies is one order of magnitude
fainter when regularisation is used during $\dfi$ estimation.

The  adaptive  aspect  of  the
regularisation allows  a powerful  smoothing of the  estimated $\dfi$  at the
edges,  and  simultaneously a  data-driven  precise  estimation  of the  quite
oscillating $\dfi$ near the center.

\subsection{Computation of $h_1$}

The $\dfi$ estimated at  $\ld_2$ is now used to compute $h_1$,  the PSF of the
first image. This  computed PSF will then be used in  the deconvolution of the
first image  $i_{\ld_1}$. Once  again we assume  to know perfectly  the static
aberrations of first imaging channel. The OTF $\tilde h_1$ is thus computed as
product    of    turbulent    OTF     and    static    OTF    by    mean    of
Equations~(\ref{eqn-sepa_h}),~(\ref{eqn-rescale}) and (\ref{eqn-otf_image2}).

\subsection{Object estimation}

\begin{figure}[h!]
\begin{center}\leavevmode
  \includegraphics[width=.6\linewidth]{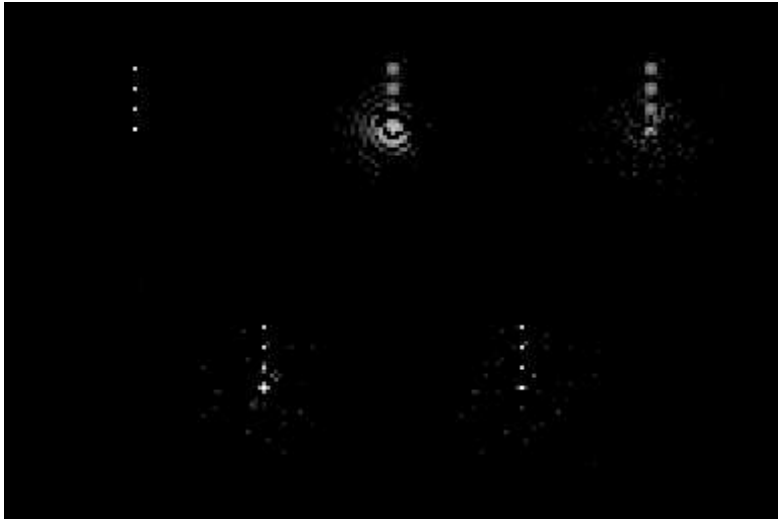}
  \caption{logarithmic scale of [Top left] observed object, and result of [Top
    middle] SD  and [Top right] DD,  [Bottom left] deconvolved  object with the
    PSF  computed  with non-regularised  structure  functions, [Bottom  right]
    deconvolved  object  with  the  PSF computed  with  regularised  structure
    function.}
  \label{fig-objets_estimes}
  \end{center}
\end{figure}

One  can  estimate  different  objects  with the  different  $\dfi$  estimated
previously : the rough $\dfi$ without regularisation, regularised $\dfi$ or by
using  the true  $\dfi$, used  for  images simulation.  The different  objects
estimated are  gathered in Figure \ref{fig-objets_estimes},  and compared with
results of differential imaging in different cases.

[Top left]  The observed object.  Central star has  a total flux  of 10$^7$
photon, the  three companions  have a ratio  of 10$^{-3}$ compared  to central
star and are situated at 2.5, 5.0 and 7.5 $\frac{\ld}{D}$ inside the AO halo.

[Top middle]  result of single differential  imaging. The  two images at
$\ld_1$ and $\ld_2$ are spatially  rescaled before subtraction. The effect of
differential aberrations on central star reduces contrast around it, the first
companion is unseeable and the second one is  visible.

[Top right] result  of double differential imaging. A  difference of reference
images  has  been  subtracted  to   the  single  difference  of  images.  This
combination  of  images  plays  the  role of  a  calibration  of  differential
aberrations and allows  to reduce their effect. This result  is ideal since it
does not  account for  slow variations of  static aberrations between  the two
images,  of evolution  of  turbulence parameters  between  the acquisition  of
object images and reference images. It is therefore a perfect DD, only limited
by photon noise.

[Bottom left] object estimated after  deconvolution by the PSF $h_{1,\dfi}$. This
PSF has  been computed with  estimated $\dfi$ with regularisation.  The two
farest companions are clearly visible, the flux ratio is almost respected. The
closest companion is  quite visible, but a bit hidden  by residual flux coming
from the central star.

[Bottom right] object estimated after deconvolution with the PSF obtained with
regularised $\dfi$. The noise in this object is slightly fainter than in the
previous one, and the central star and the first object are better defined.

\section{Performance evaluation}

The  method  performance  is  evaluated  here  by  the  5$\sigma$  detectivity
profiles. Detectivity  profile is obtained  as 5 times the  standard deviation
computed azimutally on the result of Single Difference, Double Difference, and
object estimation by our approach normalised to peak flux in image.

\begin{figure}[h!tb]
\begin{center}\leavevmode
  \includegraphics[width=.6\linewidth]{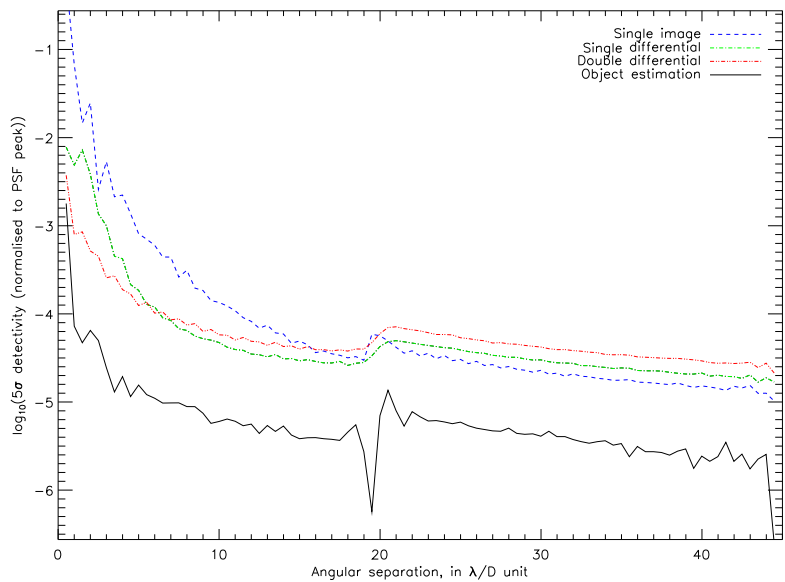}
  \caption{Averaged  detection profiles  at  $5\sigma$ in  the  case of  rough
    image,  Single difference,  double  difference and  our simplified  object
    estimation.}
  \label{fig-detection_profiles}
  \end{center}
\end{figure}

These detectivity  profiles are shown  on Figure \ref{fig-detection_profiles}.
The simulation conditions are the  one listed on section \ref{sec-results}. As
the  static  differential aberrations  are  weak  (10nm  RMS on  each  imaging
channel), the  DD is better  than SD only  close to optical axis  (closer than
5$\frac{\lambda}{D}$)  when differential  aberrations  effects dominates.  Far
from the optical axis, the two differences are photon noise limited and the SD
gives therefore  slightly better  detectivity. We found  the expected  gain in
$\sqrt{2}$.  Whatever   the  angular  separation,  our   method  gives  better
detectivity. The  gain is more than  5 in the whole  field of view.  It is due
both to the concentration of the object light in one pixel and to photon noise
reduction due to our regularised deconvolution.

\newpage
\section{Conclusion}

We  propose  a method  which  allows  to  solve SD  limitations  (differential
aberrations) and gives better results  than DD, without reference images. In a
perfect  case,  detectivity  at  $5\sigma$  reaches  less  than  $10^{-5}$  at
15$\frac{\lambda}{D}$.  This result  has still  to be  tested in  more complex
cases  (slowly  evolving   static  aberrations  or  mis-calibration,  residual
background) but  gives a rough  idea of the  potentiallity of the  method. The
perspectives  for  this method  are  to perform  a  global  estimation of  the
parameters  (structure   function,  object  in  the  two   images  and  static
aberrations), to  process real images  obtained on a differential  imager like
NACO SDI, and to generalise our approach to coronagraphy.

\bibliographystyle{spiebib}   %>>>> makes bibtex use spiebib.bst
\bibliography{Acronymes,EnglishAcronyms,Actes,Articles,Livres,Theses,BibActes,BibLivres,./fusco}

\providecommand{\inpreparationname}{en pr\'eparation}
  \providecommand{\submittedname}{soumis}
  \providecommand{\acceptedname}{accept\'e pour publication}
  \providecommand{\tobepublishedname}{\`a para\^{\i}tre}
  \providecommand{\contractname}{Contrat}
  \providecommand{\conferencedatename}{Date conf\'erence~: }
  \providecommand{\patent}[2]{Brevet #1 #2}
  \providecommand{\firstabbrevname}{1\textsuperscript{\`ere} }
  \providecommand{\secondabbrevname}{2\textsuperscript{\`eme} }
  \providecommand{\thirdabbrevname}{3\textsuperscript{\`eme} }
  \providecommand{\fourthabbrevname}{4\textsuperscript{\`eme} }
  \providecommand{\fifththabbrevname}{5\textsuperscript{\`eme} }
  \providecommand{\sixththabbrevname}{6\textsuperscript{\`eme} }
  \renewcommand{\inpreparationname}{in preparation}
  \renewcommand{\submittedname}{submitted}
  \renewcommand{\acceptedname}{accepted} \renewcommand{\tobepublishedname}{to
  be published} \renewcommand{\contractname}{Contract from}
  \renewcommand{\conferencedatename}{Conference date: }
  \renewcommand{\patent}[2]{#1 Patent #2}
\begin{thebibliography}{10}

\bibitem{Beuzit-p-05}
J.-L. Beuzit, D.~Mouillet, C.~Moutou, K.~Dohlen, P.~Puget, T.~Fusco, and
  A.~e.~a. Boccaletti, ``A planet finder instrument for the vlt,'' in {\em IAUC
  200, Direct Imaging of Exoplanets: Science \& Techniques},  2005.
\newblock \conferencedatename{}Oct. 2005, Nice, France.

\bibitem{Fusco-p-05a}
T.~Fusco, G.~Rousset, J.-L. Beuzit, D.~Mouillet, K.~Dohlen, R.~Conan, C.~Petit,
  and G.~Montagnier, ``Conceptual design of an extreme ao dedicated to
  extra-solar planet detection by the vlt-planet finder intrument,'' in {\em
  Astronomical Adaptive Optics Systems and Applications II},  ~{\bf 5903},
  Proc.\ Soc.\ Photo-Opt.\ Instrum.\ Eng., SPIE, 2005.
\newblock \conferencedatename{}July2005, San Diego, USA.

\bibitem{Boccaletti-a-04}
A.~Boccaletti, P.~Riaud, P.~Baudoz, J.~Baudrand, D.~Rouan, D.~Gratadour,
  F.~Lacombe, and A.-M. Lagrange, ``The four-quadrant phase-mask coronagraph.
  {IV}. first light at the very large telescope,'' {\em Pub.\ Astron.\ Soc.\
  Pacific}~{\bf 112}, p.~1479, 2004.

\bibitem{Marois-t-04}
C.~Marois, {\em Direct Exoplanet Imaging around Sun-like Stars: Beating the
  Speckle Noise with Innovative Imaging Techniques}.
\newblock PhD thesis, Université de Montréal, 2004.

\bibitem{Racine-a-99}
R.~Racine, G.~A. Walker, D.~Nadeau, and C.~Marois, ``Speckle noise and the
  detection of faint companions','' {\em Pub.\ Astron.\ Soc.\ Pacific}~{\bf
  112}, p.~587, 1999.

\bibitem{Boccaletti-p-05b}
A.~Boccaletti, D.~Mouillet, T.~Fusco, P.~Baudoz, C.~Cavarroc, J.-L. Beuzit,
  C.~Moutou, and K.~Dohlen, ``Analysis of ground-based differential imager
  performance,'' in {\em IAUC 200, Direct Imaging of Exoplanets: Science \&
  Techniques},  2005.
\newblock \conferencedatename{}Oct. 2005, Nice, France.

\bibitem{Sauvage-p-05}
J.-F. Sauvage, T.~Fusco, G.~Rousset, C.~Petit, A.~Blanc, and J.-L. Beuzit,
  ``Fine calibration and pre-compensation of ncpa for high performance ao
  system,'' in {\em Advancements in Adaptive Optics},  ~{\bf 5903}, pspie,
  SPIE, 2005.
\newblock \conferencedatename{}June 2005, San Diego, USA.

\bibitem{Conan-t-94}
J.-M. Conan, {\em {\'E}tude de la correction partielle en optique adaptative}.
\newblock PhD thesis, Universit{\'e} Paris XI Orsay, Oct.~1994.

\bibitem{Roddier-81a}
F.~Roddier, ``The effects of atmospherical turbulence in optical astronomy,''
  in {\em Progress in Optics},  E.~Wolf, ed., ~{\bf XIX}, pp.~281--376, North
  Holland, Amsterdam, 1981.

\bibitem{Mugnier-a-04}
L.~M. Mugnier, T.~Fusco, and J.-M. Conan, ``{MISTRAL}: a myopic edge-preserving
  image restoration method, with application to astronomical
  adaptive-optics-corrected long-exposure images.,'' {\em J.\ Opt.\ Soc.\ Am.\
  A}~{\bf 21}, pp.~1841--1854, Oct.~2004.

\end{thebibliography}

\end{document}